\documentclass[aps,pra,twocolumn,superscriptaddress]{revtex4-2}
\usepackage[english]{babel}
\usepackage{amsmath}
\usepackage{bm}
\usepackage{amssymb,amsfonts,latexsym,fancyhdr}
\usepackage{graphicx}
\usepackage {subfigure}
\usepackage[pdftex,colorlinks=true,allcolors=blue]{hyperref}
\usepackage{xcolor}
\usepackage{braket}
\usepackage{verbatim}

\newcommand{\canc}[1]{}
\newcommand{\orcidicon}[1]{\href{https://orcid.org/#1}{\includegraphics[height=\fontcharht\font`\B]{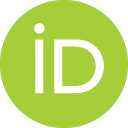}}}
\usepackage{soul}

\begin{document}
\title{Flat-band energy filtering in interacting systems: \\ conditions for improving thermoelectric performances}

\author{F. Cosco\orcidicon{0000-0002-4418-3879}}
\affiliation{Quantum algorithms and software, VTT Technical Research Centre of Finland Ltd, Tietotie 3, 02150 Espoo, Finland}
\author{R. Tuovinen\orcidicon{0000-0002-7661-1807}}
\affiliation{Department of Physics, Nanoscience Center
P.O. Box 35, 40014 University of Jyväskylä, Finland}
\author{F. Plastina}
\affiliation{Dipartimento di Fisica, Universit\`a della Calabria, 87036 Arcavacata di Rende (CS), Italy}
\affiliation{INFN, Sezione LNF, gruppo collegato di Cosenza}
\author{N.~Lo~Gullo~\orcidicon{0000-0002-8178-9570}}
\email{nicolino.logullo@unical.it}
\affiliation{Dipartimento di Fisica, Universit\`a della Calabria, 87036 Arcavacata di Rende (CS), Italy}
\affiliation{INFN, Sezione LNF, gruppo collegato di Cosenza}

\begin{abstract}
Motivated by recent theoretical and experimental studies on the role of flatbands in the thermoelectric properties of Ni$_3$In$_{1-x}$Sn$_x$ compounds, we investigate electron transport in two minimal one-dimensional flatband models, the sawtooth and diamond chains, which differ in a crucial aspect: the flatband is separated from the dispersive band by a finite gap in the former, while the two bands touch in the latter. Using a non-equilibrium Green function framework with interactions treated at the Hartree-Fock and GW levels, we compute the full set of thermoelectric coefficients and the figure of merit $zT$ as functions of gate voltage and temperature. We show that, contrary to naive expectation, a perfectly isolated flat-band is a physically ill-founded thermoelectric: the electrical conductivity vanishes as the chemical potential enters the flat-band, rendering the large Seebeck coefficient and the apparent violation of the Wiedemann-Franz law physically meaningless. Optimal thermoelectric performance is instead achieved just below the flat-band edge, where the transmission function varies most rapidly with energy, consistent with the Mahan-Sofo picture, and requires a finite broadening of the flat-band through hybridization with dispersive states. We further show that electron-electron interactions renormalize the flat-band structure itself, inducing an interaction-driven narrowing of the bandwidth and, in the diamond chain, a correlation-induced opening of a gap between the flat-band and the dispersive band near half-filling. Mean-field treatments are found to systematically overestimate \(zT\), highlighting the importance of beyond-mean-field correlations for quantitatively reliable predictions in flat-band thermoelectrics.
\end{abstract}

\maketitle

\section {Introduction}
Achieving high thermoelectric performances is currently becoming a very active research topic, especially in low dimensional materials~\cite{Hicks1993a,Hicks1993b,Dresselhaus2007,Snyder2008,Aslani2026,Zhao2016,Li2020} .
In particular understanding how electronic structure controls thermoelectric transport remains a central challenge in condensed-matter physics and materials design. In a seminal work, Mahan and Sofo ~\cite{Mahan1996} demonstrated that a delta-like transmission function optimizes the thermoelectric figure of merit \(zT\), thereby establishing energy filtering as a fundamental design principle for thermoelectric materials. This insight has motivated extensive efforts to engineer band structures and scattering mechanisms that generate narrow transport resonances.
Flat-band systems~\cite{Leykam2018} offer a particularly promising route toward realizing this principle. The dispersionless character of a flat band concentrates a macroscopic density of states within a narrow energy window, providing intrinsic spectral selectivity. However, the vanishing group velocity of perfectly flat bands suppresses direct charge transport, so that an isolated flat band contributes little to conductance~\cite{Bouzerar2021}. Recent theoretical work has shown that this limitation can be overcome when flat-band states are coupled to nearby dispersive bands: scattering and hybridization convert the large density of states into sharp resonant features in the transmission, effectively implementing the optimal energy-filter mechanism and enhancing thermoelectric response \cite{Garmroudi2025a,Garmroudi2025b}. These results highlight the importance of the relative band alignment and interband coupling in determining the transport properties of flat-band lattices.

At the same time, rapid progress in band-structure engineering has enabled the controlled realization of flat bands in low-dimensional systems. In particular, synthetic and nanoscale one-dimensional lattices have been shown to host tunable flat or nearly flat bands whose energetic position can be brought close to the Fermi level \cite{Gao2025,Tacchi2023}. Such geometries provide highly controllable minimal platforms in which the interplay between lattice topology, band flatness, and quantum transport can be studied microscopically. One-dimensional models are therefore especially well suited for identifying generic mechanisms that may extend to more complex materials.

\begin{figure}[t]
\includegraphics[width=\linewidth]{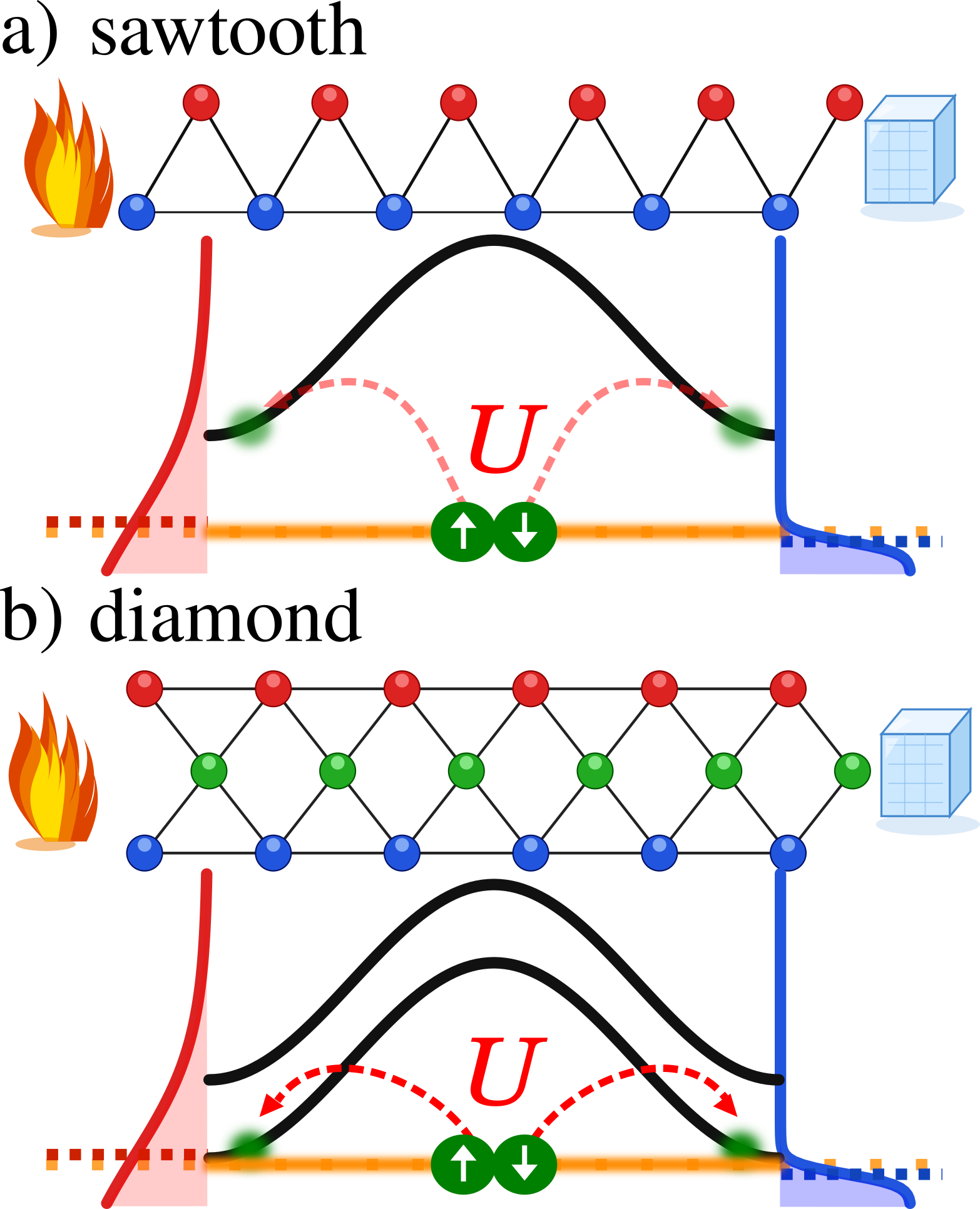}
\caption{(Color online). Schematic illustration of the two models used in this work to study the thermoelectric properties of interacting electrons in flat-band systems. Panel a) shows the sawtooth chain, which hosts an isolated flat-band separated from a dispersive band by a finite energy gap. Panel b) shows the diamond chain, whose flat-band touches the dispersive band at a single point in momentum space, with no gap. In both cases, the system is connected to a hot and a cold reservoir and electrons are subject to a local Hubbard repulsion \(U\). The Fermi-Dirac distributions of the two leads are shifted relative to one another by the temperature bias, driving a thermoelectric current through the flat-band system.
}
\label{fig:setup}
\end{figure}

In this work, we investigate the transport and thermoelectric properties of two paradigmatic one-dimensional models with flat bands: the sawtooth chain and the diamond chain~\cite{Derzhko2015,Bouzerar2022}, where the names refer to the shape of the crystal are shown in Fig.~\ref{fig:setup}.  These systems realize complementary regimes of flat-band physics. The sawtooth chain hosts an isolated flat band that is well separated from dispersive states, providing a reference limit in which flat-band modes remain largely decoupled from conducting channels. In contrast, the diamond chain features a flat-band in close proximity to dispersive bands, naturally enabling hybridization and interband scattering. This distinction allows us to directly assess how band isolation versus band mixing controls the emergence of energy-selective transmission and the resulting thermoelectric response.

Microscopically, thermoelectric transport can be addressed via density functional theory (DFT)-based approaches~\cite{Eich2017}, where charge and heat currents are treated on similar footing as the electronic ground-state problem. Within steady-state DFT, the Landauer-B{\"u}ttiker transmission can be constructed from Kohn-Sham spectra, enabling efficient evaluation of quantities such as the Seebeck coefficient and figure of merit~\cite{Sobrino2021}. Extensions to multiterminal setups~\cite{Sobrino2023} and analytically tractable model systems such as double quantum dots~\cite{Sobrino2025} have further broadened its applicability, while thermal density functional theory provides a route to include temperature effects and fluctuations~\cite{Palamara2024}. Despite this progress, DFT faces intrinsic challenges in non-equilibrium transport settings. In particular, its practical accuracy relies on approximate exchange-correlation functionals, which can be inadequate for strongly correlated systems~\cite{Burke2012}. Moreover, standard DFT is not naturally formulated for open quantum systems driven out of equilibrium, where the treatment of leads and bias voltages becomes non-trivial~\cite{DiVentra2004, Stefanucci2004, Kurth2005}. While time-dependent formulations provide a formally exact route via a partition-free setup, their practical implementation remains demanding and not straightforward for general transport scenarios~\cite{Stefanucci2007, Eich2014}. Related issues can arise when propagating scattering states~\cite{Kloss2021} or, e.g., in density-matrix renormalization group approaches~\cite{Eckel2010} requiring a full microscopic description of both the conducting device and the leads.

These limitations motivate us to use of the non-equilibrium Green function (NEGF) formalism, which provides a natural framework for open, biased quantum systems~\cite{Stefanucci2013,Ridley2022}. Within NEGF, many-body correlations are incorporated systematically through the self-energy, enabling controlled approximations beyond mean-field. This makes it particularly suitable for studying steady-state transport as well as time-dependent phenomena, including temperature-driven dynamics~\cite{Portugal2021, Pavlyukh2025, Tuovinen2025}. Here, we employ NEGF to compute charge and heat currents while retaining the full energy dependence of the transmission. Electronic interactions are treated at the level of Hartree-Fock and GW approximations, allowing us to disentangle mean-field renormalization from dynamical correlation effects. Such interactions are expected to play a central role in flat-band systems, where the enhanced density of states amplifies correlation effects and strongly reshapes spectral and transport properties~\cite{Cosco2024}.

By combining band-structure engineering with interacting quantum transport calculations, we demonstrate that proximity-induced coupling between flat and dispersive bands generates sharp transmission resonances that closely approach the ideal energy-filter condition and strongly enhance the Seebeck coefficient and thermoelectric power factor. Conversely, when the flat band is spectrally isolated, interaction effects predominantly renormalize localized states without producing significant transport enhancement. Our results elucidate how geometry, hybridization, and many-body effects cooperate to control thermoelectric performance in flat-band systems and provide design principles for low-dimensional platforms where such band structures could be engineered experimentally.

\section{NEGFs approach to thermoelectrics}
\label{sec:thermoelectric}

In Ref.~\cite{Mahan1996} the authors derived the form of the transport distribution yielding the optimal thermoelectric material. The approach relied on the analysis of the transport distribution function, \(\Sigma(\omega)\), which embodies the essential physics of both the electronic band structure and the energy-dependent scattering processes (via the electron lifetime). The key result is that the optimal thermoelectric performance is achieved when the energy transport distribution is narrowly confined, ideally approximating a delta function. Two recent studies \cite{Garmroudi2025a,Garmroudi2025b} have applied this principle to engineer the density of states in binary compounds, thereby demonstrating an enhancement in their thermoelectric power. In this section, we reframe the original derivation using the Non-Equilibrium Green Function (NEGF) formalism. This approach proves to be a powerful and natural framework for understanding the microscopic mechanisms underlying the physical conditions—specifically, a very low Lorenz ratio and a high Seebeck coefficient—that result in a high figure of merit, \(zT\).

\subsection{NEGFs approach}
Transport in quantum systems is commonly described within the Landauer–B\"uttiker framework \cite{Landauer1957,Landauer1970,Buttiker1986}, in which currents are expressed in terms of transmission coefficients through the scattering region. While this picture is exact for non-interacting systems, a realistic description of thermoelectric performance requires careful treatment of the energy dependence of the transmission function, including broadening effects arising from hybridization with the leads and electron-electron scattering. A natural generalization that incorporates these effects is the Meir–Wingreen formalism \cite{Meir1992}, in which charge and heat currents are expressed as energy integrals over a transmission function that retains the full spectral structure of the interacting system. This framework provides the foundation for the transport calculations carried out in the present work.

The NEGF formalism, particularly within the linear-response regime, offers a first-principles approach to calculating energy dependent transmission functions. For a nano-scale or microscopic conductor coupled to two electrodes (reservoirs) at slightly different temperatures and electrochemical potentials, the charge current (\(I_\alpha\)) and the energy current (\(J_\alpha\)) can be derived from the Meir-Wingreen formula. The expressions are derived in Appendix~\ref{app:negfs}. In the steady state, the expression for these currents simplify significantly and are expressed as integrals over energy (or frequency \(\omega = E/\hbar\)) of the transmission function \(\mathcal{T}(\omega)\). The latter is expressed as \(\mathcal{T}(\omega) = \text{Tr}[\Gamma_L(\omega) G^R(\omega) \Gamma_R(\omega) G^A(\omega)]\) and it is the central quantity in NEGF approach of transport. It incorporates the electronic structure through the retarded (\(G^R\)) and advanced (\(G^A\)) Green functions of the scattering region and its coupling to the leads via the broadening matrices (\(\Gamma_{L/R}\)). This inherently includes the effect of the electronic band structure and, through the inclusion of many-body self-energies, can also account for inelastic scattering processes.

The starting point for thermoelectric analysis is the expression for the electrical current (\(I_\alpha\)) and the heat current (\(\dot Q_{\alpha} = J_{\alpha} - \mu I_\alpha/e\)) flowing from the lead. Here, the chemical potential is \(\mu= -e V_g\) and \(V_g\) is the gate voltage which we set to be zero at the energy of the flat-band of the noninteracting system. In the linear-response regime, where the applied voltage difference \(\Delta V\) and temperature difference \(\Delta T\) are small, these currents can be expanded as:
\begin{equation}
\begin{pmatrix} I_\alpha \\ \dot Q_{\alpha} \end{pmatrix} = 
\begin{pmatrix} L_{11} & L_{12} \\ 
L_{21} & L_{22} \end{pmatrix} 
\begin{pmatrix} \Delta V \\ \Delta T \end{pmatrix}
\end{equation}
The above are also called the Onsager reciprocal relations. The Onsager coefficients \(L^{ij}\) can be directly derived from the NEGF expression for the current~\ref{app:negfs}. By comparing these derived coefficients with those used in Ref.~\cite{Mahan1996}, we find a direct correspondence. The kinetic coefficients defined in Eqs.~(2)--(4) of Ref.~\cite{Mahan1996} are:
\begin{align}
    L_{11} &= e^2  \mathcal{I}_0,  \hspace{1 cm} L_{22} = \frac{1}{T} \mathcal{I}_2 \\
    L_{12} &= \frac{e}{T}  \mathcal{I}_1 ,  \hspace{1 cm} L_{21} = L_{12}T 
\end{align}
where the integrals \( \mathcal{I}_n\) are defined as:
\begin{equation}
 \mathcal{I}_n = \frac{1}{\hbar} \int_{-\infty}^{\infty} \frac{d\omega}{2\pi} \left(-\frac{\partial f}{\partial \omega}\right) \mathcal{T}(\omega) (\omega-\mu)^n
\end{equation}
with \(f(\omega)=[e^{\beta(\omega-\mu)}+1]^{-1}\) being the Fermi-Dirac distribution function with $\beta\equiv (k_B T)^{-1}$. 
The electrical conductivity \(\sigma\), the Seebeck coefficient \(S\), and the electronic thermal conductivity \(\kappa_e\) are then given by (see Appendix~\ref{app:negfs}):
\begin{align}
    \sigma &= e^2  \mathcal{I}_0 \label{eq:sigma} \\
    S &= \frac{1}{eT} \frac{ \mathcal{I}_1}{ \mathcal{I}_0}=e \frac{\mathcal{I}_1}{\sigma T} \label{eq:seebeck} \\
    \kappa_e &= \frac{1}{T} \left(  \mathcal{I}_2 - \frac{ \mathcal{I}_1^2}{ \mathcal{I}_0} \right) \label{eq:kappae}
\end{align}
These expressions [Eqs.~\eqref{eq:sigma}--\eqref{eq:kappae}] are identical in form to those derived in Ref.~\cite{Mahan1996}. The key equivalence is that the transport distribution function \(\Sigma(\omega)\) in their work is replaced by \(\mathcal{T}(\omega)/(2\pi\hbar)\). 

\subsection{The Condition for Optimal Performance}

The thermoelectric figure of merit, \(zT\), which quantifies the thermoelectric performances can be rewritten as:
\begin{equation}
zT = \frac{\sigma S^2 T}{\kappa_e + \kappa_{ph}} = \frac{S^2}{L + L_{ph}}
\label{eq:zt}
\end{equation}
where \(\kappa_{ph}\) is the phonon thermal conductivity, \(L = \kappa_e/(\sigma T)\) is the electronic Lorenz ratio, and \(L_{ph} = \kappa_{ph}/(\sigma T)\). The search for the optimal shape of \(\Sigma(\omega)\) yielding the highest possible \(zT\) can be reformulated as a mathematical problem of finding the distribution function \(P(\omega)=-\partial f/\partial \omega \Sigma(\omega)\) whose moments are the \(I_n\) and such that the \(zT\) in Eq.~\eqref{eq:zt} is maximal. The solution to this variational problem is a Dirac delta function, \(\Sigma(\omega) \propto \delta(\omega - \mu - \epsilon_0)\), centered at an optimal energy \(\epsilon_0\) away from the chemical potential \(\mu\).

Our analysis, grounded in the NEGF expressions above, allows us to understand the physical rationale behind this mathematical result. The conditions derived in Ref.~\cite{Mahan1996} for a high \(zT\) correspond to a very low Lorenz ratio \(L\) and a large Seebeck coefficient \(S\) (respectively  (\(\xi \rightarrow 1\) and \(A\rightarrow 0\) in Ref.~\cite{Mahan1996}). 
Let us analyze the two main players which need to be tuned in order to achieve the highest possible \(zT\).

\textbf{Lorenz Ratio (\(L\)):} The Lorenz ratio \(L = \kappa_e/(\sigma T)\) is a measure of the electronic thermal conductivity relative to the electrical conductivity. Using the definitions above, we can write:
\begin{equation}
L = \frac{1}{e^2 T^2} \left( \frac{\mathcal{I}_2}{\mathcal{I}_0} - \left(\frac{\mathcal{I}_1}{\mathcal{I}_0}\right)^2 \right).
\end{equation}
By noticing that \(\mathcal{I}_0\) is the normalization of the distribution \(P(\omega)\), this expression is essentially the variance of the energy of the transmitted electrons. There are two distributions which satisfy this condition: i) the rectangular distribution; ii) the delta-function. The derivative of the Fermi-Dirac distribution is peaked around the chemical potential \(\mu\) and has a width \(\approx k_B T\). Therefore a delta-function distribution \(P(\omega)\) is possible for a sharply peaked transmission function \(\mathcal{T}(\omega) \propto \delta(\omega - \omega_0)\). In this case all transmitted electrons have exactly the same energy \(\omega_0\). Consequently, the variance is zero, and we get \(\mathcal{I}_2/\mathcal{I}_0 = (\mathcal{I}_1/\mathcal{I}_0)^2\), leading to \(\kappa_e = 0\) and thus \(L=0\). This is the "best" possible scenario, as it completely decouples charge and heat transport electronically. In reality, a delta function is nonphysical, but a very sharp, narrow-band transmission function approximates this behavior, resulting in a very small \(L\). At high temperatures where the Fermi window broadens significantly, a rectangular distribution would also lead to a low Lorenz factor.
    
\textbf{Seebeck Coefficient (\(S\)):} From Eq.~\eqref{eq:seebeck}, the Seebeck coefficient is proportional to the ratio between the average energy (relative to the chemical potential) of the transported electrons and the energy due to electrons with thermal energy \(k_B T\). In the case of a rectangular distribution, centered around the chemical potential \(\mu\), the integral \(\mathcal{I}_1\approx 0\), leading to a vanishing \(S\). This is in general true for any distribution which is symmetric around \(\mu\). Instead, \(S\) is maximized when the transport is asymmetric around \(\mu\). Being the Fermi window (\(-\partial f/\partial \omega\)) symmetric around \(\mu\), this occurs when \(\mathcal{T}(\omega)\) is significant only on one side of \(\mu\). A delta-function transmission located at an energy \(\epsilon_0\) away from \(\mu\) achieves this purpose. It ensures that all charge carriers have exactly the same energy, maximizing the \(\mathcal{I}_1/\mathcal{I}_0\) ratio and, consequently, \(S\).

Therefore, the condition for an exceptionally high \(zT\) is to have a transport distribution (or transmission function) that is both extremely narrow (for a low \(L\)) and asymmetric (for a high \(S\)). The \(zT\) then becomes \(zT = S^2 / L_{ph}\), where the denominator is now dominated by the unavoidable phonon thermal conductivity and which bounds it from above. A large \(S\) can then compensate for a finite phononic contributions \(L_{ph}\) and significantly enhance \(zT\).

Our subsequent analysis of specific models, such as the sawtooth lattice and the diamond chain, directly applies this principle. While both models may exhibit a delta-like density of states from a flat band, it is the transmission function \(\mathcal{T}(\omega)\) that dictates transport. In the diamond chain, the topology allows for a non-zero transmission at the flat-band energy, whereas in the sawtooth model, it may vanish. This highlights the crucial role of the quantum-mechanical transmission probability, accessible via NEGF, over the simple density of states in the quest for high-performance thermoelectrics. The transmission function, and not merely the density of states, is the true design parameter.

Motivated by the recent realization of one-dimensional (1D) \cite{Gao2025} and quasi-1D \cite{Wagner2023,Bouaziz2024} flat-band we consider two well known 1D models. The simplest one is the sawtooth  model, a one-dimensional lattice with a unit cell hosting two different sites (or orbitals) \(a\) and \(b\) coupled as shown in Fig.\ref{fig:setup} a). The second model we will consider is the so-called diamond model which in addition to a flat-band can host topological non-trivial phases. Its unit cell hosts three different sites (or orbitals) \(a\), \(b\) and \(c\) which are coupled as in Fig.\ref{fig:setup} b). 
We notice that in both models the interaction is up to the "nearest-neighbor cell". We can write a general expression for the Hamiltonian which holds for both models as:
\begin{equation}
\label{eq:MHam}
\hat H_0=\sum\limits_{i=1,\sigma}^{N} \hat {\mathbf c}_{i\sigma}^{\dagger}\textbf{\text{h}}\; \hat {\mathbf c}_{i\sigma}+\sum\limits_{i=1,\sigma}^{N-1} \hat {\mathbf c}_{i+1\sigma}^{\dagger}\textbf{\text{T}} \hat {\mathbf c}_{i\sigma}+\hat {\mathbf c}_{i\sigma}^{\dagger}\textbf{\text{T}}^\dagger \hat{\mathbf c}_{i+1\sigma},
\end{equation}
where $N$ is the number of cells.
The two models are then obtained by properly choosing the intra-cell \(\textbf{\text{h}}\) and inter-cell \(\textbf{\text{T}}\) matrices and the vector operators \({\hat{\mathbf c}}_{i\sigma}\equiv \{\hat{c}_{i\alpha\sigma}\}\) and \(\hat{\mathbf c}_{i\sigma}^\dagger\equiv\{\hat{c}_{i\alpha\sigma}^{\dagger}\}^\text{T}\) with \(\alpha\) running over the orbitals in the unit cell.

In what follows we work with periodic boundary conditions in order to avoid spurious effects coming from localized edge states which are in gap. To get rid of such effects in a recent work ~\cite{Pyykkonen2021,Pyykkonen2023} the authors introduced a local potential on one of the edge sites to push the localized state within the band of states. Since we do not look at topological properties, these states do not play any crucial role in the forthcoming discussion. Moreover in a realistic setup the contacts with the leads would destroy the boundary states because of the hybridization with the atoms of the lead itself.
With this working assumption, it is possible to diagonalize~\cite{Stefanucci2013} the Hamiltonian in Eq.~\eqref{eq:MHam} introducing the lattice momentum obtaining:
\begin{equation}
\label{eq:MHamk}
\hat H_0=\sum\limits_{k=-N/2+1,\sigma}^{N/2} \tilde {\mathbf c}_{k\sigma}^{\dagger}\widetilde{\textbf{\text{h}}}(k)\; \tilde {\mathbf c}_{k\sigma}.
\end{equation}
where \(\widetilde{\textbf{\text{h}}}(k)= \textbf{\text{h}}+\exp(ik)\textbf{\text{T}}+\exp(-ik)\textbf{\text{T}}^\dagger\) and \(\tilde {\mathbf c}_{k\sigma}=N^{-1/2}\sum\limits_{j=1}^{N}\hat {\mathbf c}_{j\sigma}e^{-i kj }\).

In the following we briefly discuss the salient features of the two models, highlighting those which will be used in the forthcoming discussions.

\subsection{The sawtooth model}
The sawtooth model is the simplest one-dimensional model featuring a flat-band. It is defined by the matrices:
\begin{equation}
\label{eq:STmatrices}
\textbf{\text{h}}=
\begin{pmatrix}
\epsilon_a & t_{ab} \\
t_{ab} & \epsilon_b 
\end{pmatrix},\;\;\;
\textbf{\text{T}}=
\begin{pmatrix}
t_{aa} & t_{ab} \\
0 & 0
\end{pmatrix}.
\end{equation}
The operators are \({\hat{\mathbf c}}_{i\sigma}\equiv \{\hat{c}_{i\alpha\sigma}\}\) \({\hat{\mathbf c}}_{i\sigma}^{\dagger}\equiv \{\hat{c}_{i\alpha\sigma}^{\dagger}\}^\text{T}\) with \(\alpha=a,b\) being two orbitals per unit-cell.

A real system described by an Hamiltonian similar to the sawtooth, the zig-zag model, has recently been realized using Cu and Te atoms on Cu(111) as reported in Ref.~\cite{Gao2025}. There it has been shown that the chain of CuTe atoms induces a flat-band at about 1.5 eV below the Fermi energy. It has been shown that its  presence induces a temperature dependent phase transition and induces a Luttinger liquid behavior.

For such a simple system it is possible to obtain analytic expressions for the dispersion relations of the two bands:
\begin{equation}
\label{eq:STeigvals}
\epsilon_{\pm}(k)=\frac{\epsilon_a+\epsilon_b}{2}+t_{aa}\cos{(k)}\pm\frac{1}{2}\sqrt{\Delta(k)},
\end{equation} 
where \(\Delta(k)=(\epsilon_a-\epsilon_b+2t_{aa}\cos{(k)})^2+8t_{ab}^2(1+\cos{(k)})\) and \(k=2m\pi/N\) with \(m\in(-N/2,N/2]\).

As shown in Fig.~\ref{fig:setup} a), the spectrum has a gap \(E_g=\epsilon_+(\pi)-\epsilon_-(\pi)=|\epsilon_a-\epsilon_b-2t_{aa}|\) which closes for \(\epsilon_a-\epsilon_b=2t_{aa}\).
To guarantee the existence of a flat-band, it is sufficient to impose \(\Delta(k)=4(C+t_{aa}\cos(k))^2\), with \(C\) a constant to be determined.
After some simple algebra, the condition for the existence of the flat-band reads 
\begin{equation}
\label{eq:flbnd_cond}
t_{ab}=\pm\sqrt{2 t_{aa}^2+(\epsilon_a-\epsilon_b)t_{aa}}.
\end{equation}

In the case of homogeneous on-site potentials \(\epsilon_a=\epsilon_b\) this condition simplifies to \(t_{ab}=\sqrt{2}t_{aa}\).
For our purpuses it is important to notice that when a perfect flat band is present the system has always a gap. In other words in the sawtooth model the flat-band, when present, is always isolated from the dispersive band.
Moreover Eq.~\eqref{eq:STeigvals} shows that the appearance of a flat-band is rather fragile with respect to the variation of the parameters of the Hamiltonian. The flat-band becomes dispersive as soon as Eq.~\eqref{eq:flbnd_cond} is not satisfied.
What is more important is that the number of states in the flat-band is an extensive quantity, and grows with the number of unit cells, namely the flat-band is a highly degenerate manifold of the system.

\subsection{The diamond chain}
The second model we consider is the diamond chain characterized by the matrices:
\begin{equation}
\label{eq:DMmatrices}
\textbf{\text{h}}=
\begin{pmatrix}
\epsilon_a & t & 0\\
t & \epsilon_b & t\\
0 & t & \epsilon_c\\
\end{pmatrix},\;\;\;
\textbf{\text{T}}=
\begin{pmatrix}
t' & t & 0\\
0 & 0 & t \\
0 & 0 & t' \\
\end{pmatrix}.
\end{equation}
The operators are \({\hat{\mathbf c}}_{i\sigma}\equiv \{\hat{c}_{i\alpha\sigma}\}\) and \({\hat{\mathbf c}}_{i\sigma}^{\dagger}\equiv \{\hat{c}_{i\alpha\sigma}\}^{\dagger}\) with \(\alpha=a,b,c\) being the site (orbital) labels.

When the local energies \(\epsilon_\alpha\) are all the same,
the diamond chain features a flat-band either for \(t'=0\) or for \(t'=t\)~\cite{Kobayashi2016}. As illustrated in Fig.~\ref{fig:setup} b) , the two cases differ in two main aspects, both important for our purposes. The flat-band at \(t'=0\) is at zero energy and three bands touch at the edge of the Brillouin zone: the system is gapless. At \(t=t'\) the flat-band is the lowest (or highest depending on the sign of \(t\)) energy band with the other two bands having a semi-metallic character. Most importantly there is no energy gap separating the flat-band and the middle energy band in the case \(\epsilon_a=\epsilon_b=\epsilon_c\) and \(t=t'\). 
This is the ideal scenario to study how transport properties of the system are modified by the coupling of the flat-band to other bands.
In the following we will consider the approach to the flat-band by tuning the local energies of orbitals \(a\) and \(c\), namely \(\epsilon_a=-\epsilon_c=\epsilon\rightarrow 0\)

\begin{figure}[t]
\includegraphics[width=\linewidth]{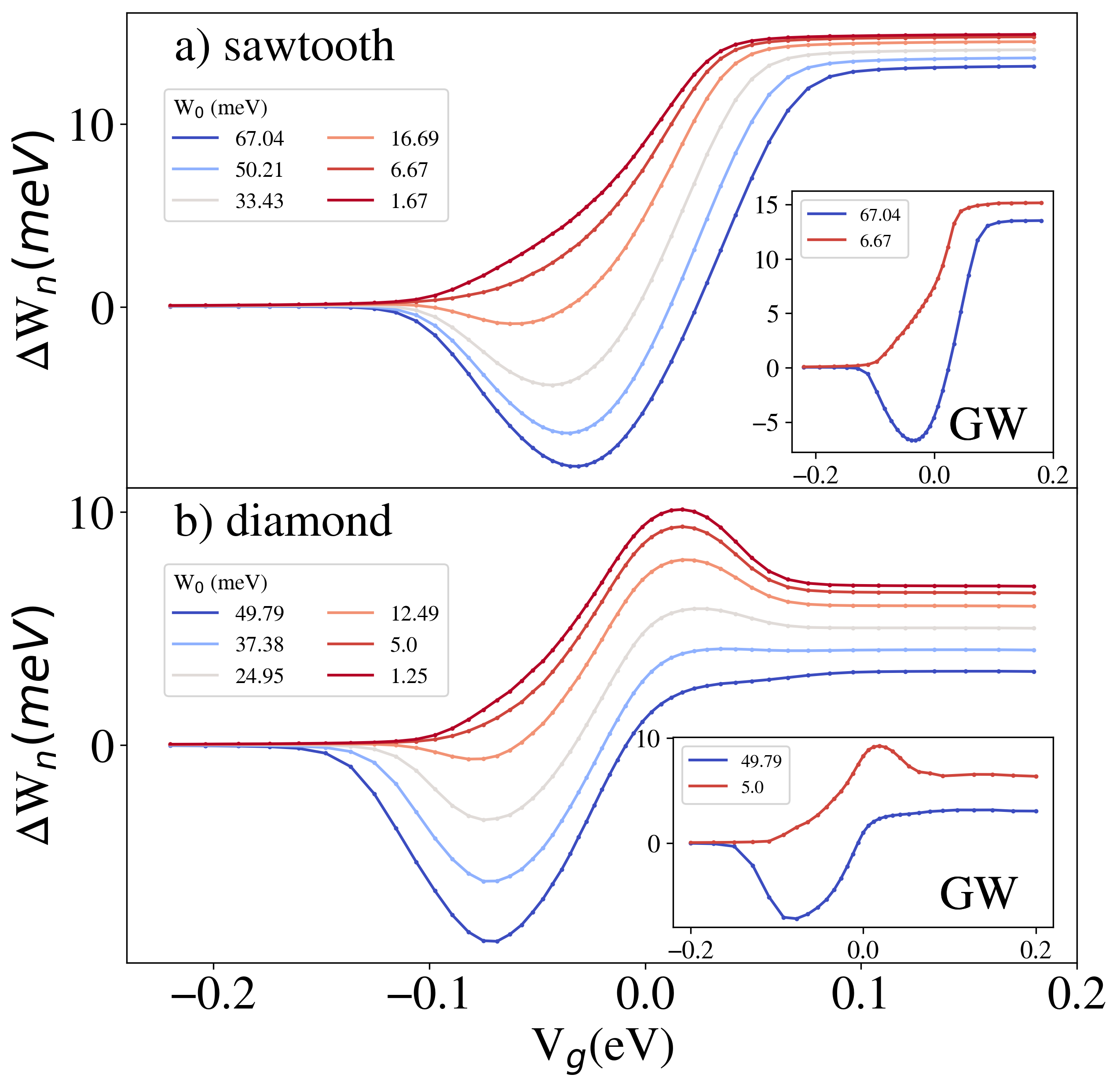}
\caption{(Color online). Interaction-induced renormalization of the narrow-band bandwidth $\Delta W_n$ as a function of the gate potential $V_g$, computed at the Hartree-Fock level for an interaction strength $U = 0.15$ eV, for panel a) the sawtooth chain and panel b) the diamond chain. Each curve corresponds to a different value of the bare lead bandwidth $W_0$ (in meV), ranging from the wider bands (blue) to the narrower bands (dark red), as indicated in the legend. The insets show the corresponding GW results for two representative values of $W_0$.}
\label{fig:bandwidths}
\end{figure}

\subsection{Interaction and contacts}

States in the flat-band do not contribute to transport directly; yet when they are coupled to dispersive states of the spectrum the system conduction is restored~\cite{Garmroudi2025a,Garmroudi2025b}. This coupling can be provided by external fields (photo-excitation), electron-electron or electron-phonon scattering or by the contact with metallic leads in the case of transport setups. 
Here we examine the role of electron-electron scattering and the contact to external leads. The latter ones are modeled as non-interacting electrons with a given dispersion relation at equilibrium and characterized by a chemical potential and a temperature. For the many-body interaction we consider a Hubbard like interaction. This is partly motivated by the original works which introduced the Hubbard model itself~\cite{Hubbard1963,Hubbard1964} to describe localized electrons in \(d-\) and \(f-\) orbitals. Nevertheless here we are considering the case in which the many-body interaction is not the leading term in the total energy as in the case of the Hubbard model. In this situation, the flatbands emerge due to geometrical arrangements of the Hamiltonian rather than from strong interactions favoring the occupation of a degenerate manifold of atomic like orbitals.

\begin{figure}[t]
\includegraphics[width=\linewidth]{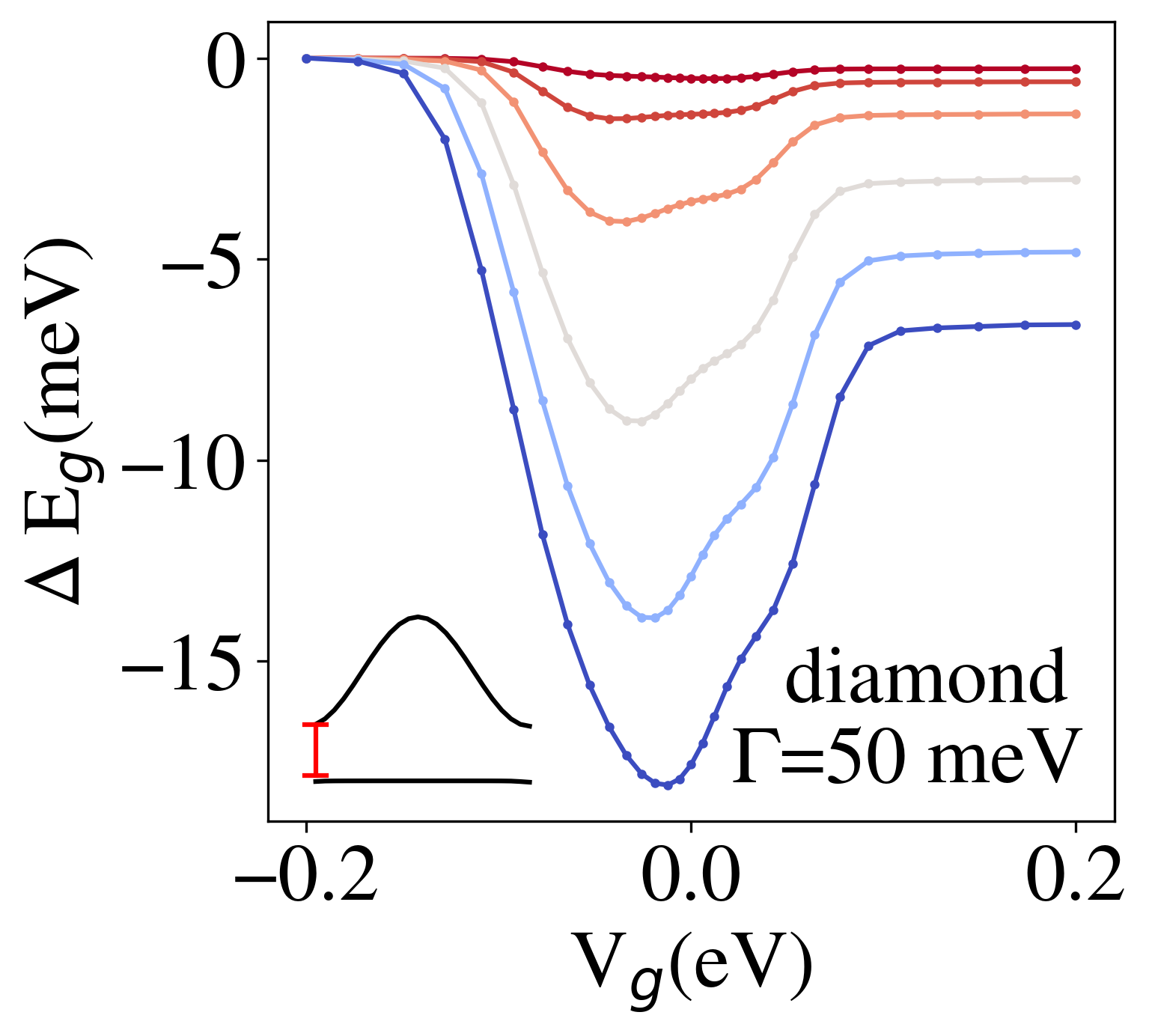}
\caption{(Color online). Interaction-induced renormalization of the energy gap \(\Delta E_g\) between the narrow-band and the dispersive band in the diamond chain, as a function of the gate potential \(V_g\), computed at the Hartree-Fock level for \(\Gamma = 50\) meV, as illustrated in the inset. Each curve corresponds to a different bandwidth \(W_0\), ranging from wider (blue) to narrower (dark red) as in Fig.~\ref{fig:bandwidths}. The renormalization is largest near half-filling of the narrow-band (\(V_g \approx 0.0\) eV).}
\label{fig:gap}
\end{figure}

The total Hamiltonian reads:
\begin{eqnarray}
&\hat H =&\hat H_{M}+\hat{H}_{ee}+\sum\limits_{l=L,R}\hat H_l+\hat V_l\\
\label {eq:ham-flat}
&\hat{H}_{ee}=& U \sum_{\substack{i=1\\\alpha}}^{N} \hat c^\dagger_{i \alpha\uparrow  } \hat c^\dagger_{i\alpha \downarrow} \hat c_{i \alpha\downarrow  } \hat c_{i\alpha \uparrow }\\
&\hat H_l =& \sum_{k, \sigma=\uparrow\downarrow} \epsilon_{l,k}  \hat d^\dagger_{l, k \sigma} \hat d_{l, k\sigma}\\
&\hat V_l =& \sum_{\substack{i=1,k \\ \sigma=\uparrow\downarrow}} \left[T_{i\alpha k}^{(l)} \hat c^\dagger_{i\alpha \sigma } \hat d_{l,k \sigma}+ T_{i\alpha k}^{(l)*} \hat d^\dagger_{{l, k \sigma }} \hat c_{i\alpha \sigma }\right],
\end{eqnarray}
where \(\hat d_{{l, k \sigma }},\hat d^\dagger_{{l, k \sigma }}\) are the annihilation and creation operators for the \(k-\)state of an electron in the lead \(l\).
Here the coefficients \(T_{i\alpha k}^{(l)}\) are the overlap integrals between the \(k-\)state of the lead \(l\) and the site (orbital) \(\alpha\) in the unit cell \(i\).

We resort to the non-equilibrium Green functions approach to compute the physical quantity of interest. We give a brief overview of the formalism in Appendix~\ref{app:negfs} with particular emphasis on the calculation of the quantities useful for the our purposes.
We mention here that in the NEGFs formalism the coupling to the leads is accounted for in non-perturbative manner, provided one knows the density of states of the leads and the overlap integrals \(T_{i\alpha k}^{(l)}\). 
This is done via the lead-induced hybridization self-energies, which we evaluate in the wide-band limit, where the electrode density of states is assumed to be featureless around the Fermi energy. This approximation is commonly justified for metallic contacts such as gold, whose electronic structure varies only weakly on the energy scales relevant for transport \cite{Viljas2005,Verzilj2013,Covito2018}.

On the other hand electron-electron interactions are described via a chosen approximation scheme. In this work we consider two different cases: the mean-field (Hartree) approximation and the self-consistent GW approximation, which accounts for dynamical screening of the interaction through polarization processes beyond the single-insertion level. The GW approximation captures important correlation effects in both molecular and extended systems, while providing a reasonable description of spectral properties in many cases~\cite{Golze2019}. In the case of a short range many-body interaction, such as the one considered in this work, the GW approximation results in an effective long range correlation built up on the single particle coherences.

\subsection{Band flattening}
\label{subsec:flattening}
In Refs.~\cite{Xie2020,Lewandowski2021,Choi2021,Dale2023} it has been shown that interactions can favor band flattening, a mechanism ascribed to the Hartree potential generated by electrons added to the system upon changing its filling. To reproduce this scenario we couple our system to an additional lead acting as a gate, which controls the total number of particles by fixing the chemical potential at \(\mu_G=e\;V_g\). At stationarity, no electrical or energy current flows through the system; instead, the system reaches equilibrium at a filling determined by the gate voltage. This corresponds to the system being in contact with a metallic Fermi sea.

We focus on the evolution of the lowest energy band as the system is progressively filled, and in particular on its bandwidth. In the non-interacting case we denote by \(W_n^0\) the bandwidth of the narrowest band, where the subscript \(n\) stands for "narrow". In the interacting case the problem is in general not reducible to a single-particle one; nevertheless, a single-particle spectrum can still be defined through the mean-field Hamiltonian. The latter contains only the Hartree term, and has the same structure as Eq.~\eqref{eq:MHamk} with the replacement \(\epsilon_\alpha \rightarrow \epsilon_\alpha + U n_{i\alpha}\) in the diagonal elements of \({\bf h}\), where \(n_{i\alpha} = \int d\omega/(2\pi)\, G_{(i,\alpha,\uparrow)(i,\alpha,\uparrow)}^{<}(\omega)\) is the occupation at site \(i\) and orbital \(\alpha\), independent of spin due to spin symmetry (see Appendix~\ref{app:negfs}).
To quantify the interaction-induced flattening we study the bandwidth change \(\Delta W_n\), with \(\Delta W_n = W_n - W_n^0\), as a function of the gate chemical potential and for different values of the bare non-interacting bandwidth \(W_n^0\). The latter in the case of the sawtooth model is given by \(W_n^0=\epsilon_+(\pi)-\epsilon_-(\pi)\) and it is defined similarly for the mean-field Hamiltonian. The bandwidth \(W_n\) of the interacting system is computed from the mean-field Hamiltonian of the contacted system. The results are shown in Fig.~\ref{fig:bandwidths} for both models with \(N=24\) and for \(U = 0.15\) eV, \(T \approx 80\) K, lead coupling \(\Gamma = 50\) meV, and gate coupling \(\Gamma_g = 1\) meV.

For both the sawtooth and diamond chains, the narrowband of the interacting system becomes flatter than its non-interacting counterpart over a wide range of gate voltages, indicating that the flattening mechanism does not depend on whether the narrowband is isolated from the rest of the spectrum. The effect is most pronounced for intermediate values of \(W_n^0\), as shown by the curves ranging from wide-band (blue) to narrow-band (dark red) limits in Fig.~\ref{fig:bandwidths}. A closer inspection reveals an important exception: when the non-interacting bandwidth is already very small, the band instead becomes more dispersive upon interaction. In this limit the system's occupation changes abruptly as the gate voltage is swept across the narrow-band, and the Hartree potential acts to broaden rather than flatten the band.

This behavior can be traced back to the structure of the non-interacting eigenstates. In the sawtooth model, the lowest-energy eigenstates have higher amplitude on the sites with lower local chemical potential $\epsilon_b$, and are therefore filled first as the gate voltage is increased. The resulting increase in local occupation raises the on-site energy via the Hartree potential, effectively softening the condition for flatband formation in Eq.~\eqref{eq:flbnd_cond}. Upon further filling, eigenstates with higher amplitude on the "A" orbitals begin to be populated, leading, again through the repulsive interaction, to a violation of the flatband condition and a subsequent increase in dispersion. An analogous picture holds for the diamond chain, although in that case a simple argument based on on-site energies alone does not apply, since \(\epsilon_c > \epsilon_b > \epsilon_a\) and yet the lowest-energy states have higher amplitude on orbitals "A" and "C".

The band flattening is a predominantly mean-field effect. As shown in the insets of Fig.~\ref{fig:bandwidths} for two representative values of \(W_n^0\), going beyond mean field to the GW level does not qualitatively alter the picture but leads to quantitative corrections, particularly near half-filling where correlation effects are strongest. It is also worth noting that any observed flattening remains small compared to the non-interacting bandwidth \(W_n^0\), while the transition to a more dispersive regime can produce bandwidth increases that significantly exceed the non-interacting value. Finally, regarding the gap between the narrow-band and the dispersive band in the diamond chain we computed the difference \(\Delta  E_g= E_g-E_g^0\) between the gap of the interacting system \(E_g\) and the gap of the non-interacting one \(E_g^0\). Fig.~\ref{fig:gap} shows that interactions drive a purely correlation-induced closing of a gap between the two bands, which is largest near half-filling and is suppressed as \(W_n^0\) increases, consistent with the picture that the effect weakens as the flat-band becomes less isolated from the dispersive background.

\section{High thermoelectric power in narrow-band systems}

\begin{figure}[t]
\includegraphics[width=\linewidth]{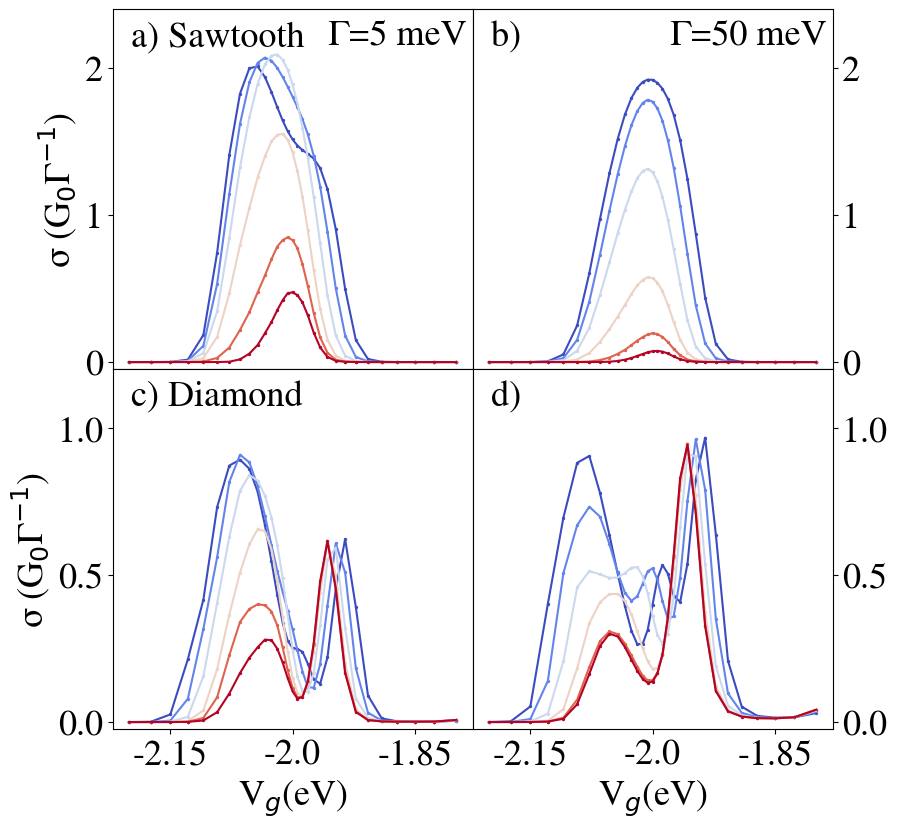}
\caption{(Color online). Electrical conductivity \(\sigma\) in units of \(G_0 \Gamma^{-1}\) as a function of the gate potential \(V_g\), computed at the Hartree-Fock level for interaction strength \(U = 0.15\) eV, temperature \(T = 80\) K, and two values of the lead broadening \(\Gamma = 5\) meV (left panels) and \(\Gamma = 50\) meV (right panels), for the sawtooth chain (panels a), b)) and the diamond chain (panels c), d)). Each curve corresponds to a different value of the bare non-interacting bandwidth \(W_n^0\), ranging from the wide-band limit (blue) to the narrow-band limit (dark red), with the same values as in Fig.~\ref{fig:bandwidths}. The insets show the corresponding GW results for two representative values of\(W_n^0\), demonstrating that correlations beyond mean field primarily renormalize the flatband-associated feature while leaving the dispersive contribution qualitatively unchanged.}
\label{fig:eq_el_conductivity}
\end{figure}

\subsection{Electric and thermal conductivities}
\label{subsec:eq_transport}

\begin{figure}[t]
\includegraphics[width=\linewidth]{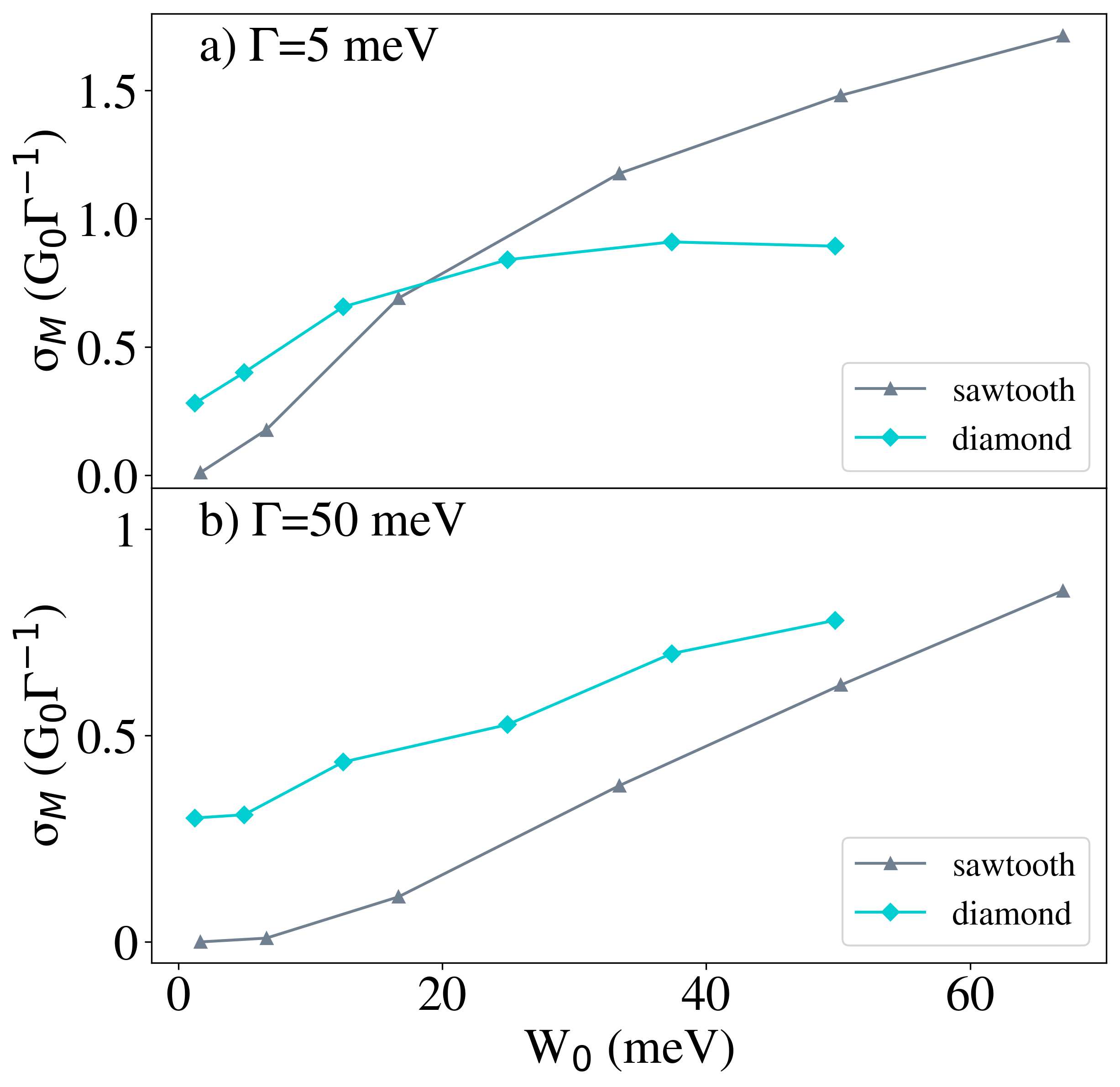}
\caption{(Color online). Maximum electrical conductivity \(\sigma_M\) in units of \(G_0\Gamma^{-1}\) as a function of the bare non-interacting bandwidth \(W_0\), computed at the Hartree-Fock level for \(U = 0.15\) eV and \(T = 80\) K, for the sawtooth (gray triangles) and diamond (cyan diamonds) chains, and for two values of the lead broadening \(\Gamma = 5\) meV (panel a) and \(\Gamma = 50\) meV (panel b). }
\label{fig:max_eq_el_conductivity}
\end{figure}
Before discussing the thermoelectric properties, we examine the electrical and thermal conductivities and compare them with the expectations from the non-interacting case. Within linear response theory, the electrical conductivity is expected to vanish in the flatband limit. This behavior results from the complete suppression of kinetic energy despite the presence of a formally divergent density of states: although infinitely many states are available, none of them contribute to transport. This conclusion, however, holds strictly only when the flatband is isolated from other energy bands.

Fig.~\ref{fig:eq_el_conductivity} shows the electrical conductivity in units of \(G_0\Gamma^{-1}\), where \(G_0 = 2e^2/h\) is the conductance quantum, as a function of the gate potential \(V_g\), computed at the Hartree-Fock level for \(U = 0.15\) eV and \(T = 80\) K, for two values of the lead broadening \(\Gamma = 5\) meV and \(\Gamma = 50\) meV and for both the sawtooth (panels a, b) and diamond (panels c, d) chains. Each curve corresponds to a different value of the bare non-interacting bandwidth \(W_n^0\), ranging from the wide-band (blue) to the narrow-band (dark red) limit. In all panels the conductivity is suppressed as \(W_n^0\) is reduced, reflecting the progressive localization of carriers as the flatband becomes more isolated. The sawtooth chain displays a single broad peak associated with transport through the dispersive band, which narrows and shifts toward lower gate voltages with decreasing \(W_n^0\). The diamond chain shows a richer two-peak structure, with a broad feature from the dispersive band and a sharp, narrower peak associated with transport through the flatband, whose weight and position depend sensitively on both \(W_n^0\) and \(\Gamma\).

The behavior in the flatband limit depends critically on both the coupling strength and the band topology, as made explicit in Fig.~\ref{fig:max_eq_el_conductivity}, which shows the peak conductivity \(\sigma_M\) as a function of \(W_0\) for both models. At small coupling (\(\Gamma = 5\) meV), both models recover the non-interacting expectation: \(\sigma_M\) increases monotonically with \(W_0\) and vanishes as \(W_0 \to 0\), consistent with the progressive suppression of transport as the flatband becomes more isolated. At larger coupling (\(\Gamma = 50\) meV) the two models diverge qualitatively. In the sawtooth chain \(\sigma_M\) continues to vanish in the flatband limit, a direct consequence of the large gap \(\Delta E_g \approx 2\) eV separating the narrowband from the dispersive band, which prevents any hybridization-driven restoration of transport regardless of the coupling strength. In the diamond chain, by contrast, \(\sigma_M\) saturates to a finite value as \(W_0 \to 0\), demonstrating that the gapless touching between the flatband and the dispersive band enables a residual transport channel that persists even when the bandwidth is fully quenched.

This observation leads to an important conclusion: transport through a flatband is restored when the latter is energetically adjacent to a dispersive band and a coupling mechanism between the two exists. This is precisely the energy filtering mechanism invoked in Refs.~\cite{Garmroudi2025a,Garmroudi2025b} to enhance thermoelectric performance through the energy selectivity of a narrowband embedded in a dispersive background. In the present case the coupling is provided by hybridization with the leads, but an entirely analogous role can be played by electron-electron interactions. When correlation effects are incorporated at the GW level, shown in the insets of Fig.~\ref{fig:eq_el_conductivity} for two representative values of \(W_n^0\), the conductance peak associated with the narrow-band persists in the diamond chain while it is suppressed in the sawtooth model, confirming that the restoration of transport is governed by the band topology rather than by the details of the interaction, and is controlled by the presence or absence of a gap between the flat-band and the dispersive background.

The thermal conductivity displays an analogous behavior to the electrical one, as shown in panels a) and c) of Fig.~\ref{fig:eq_th_conductivity} where we report \(\kappa_e\) in units of the quantum of thermal conductance \(g_0 = \pi^2 k_B^2 T(3h)^{-1}\). In the sawtooth chain, \(\kappa_e\) is progressively suppressed as \(W_n^0\) is reduced and vanishes in the flatband limit, mirroring the behavior of electrical conductivity \(\sigma\). In the diamond chain, by contrast, \(\kappa_e\) retains a finite value even as \(W_n^0 \to 0\), consistent with the restoration of transport through the gapless touching between the flatband and the dispersive band. In both models the thermal conductivity develops a two-peak structure in the diamond chain that directly reflects the two transport channels identified in the electrical conductivity, while the sawtooth chain shows a single broad peak that narrows and is suppressed with decreasing \(W_n^0\).

\subsection{Violation of the WF law}

\begin{figure}[t]
\includegraphics[width=\linewidth]{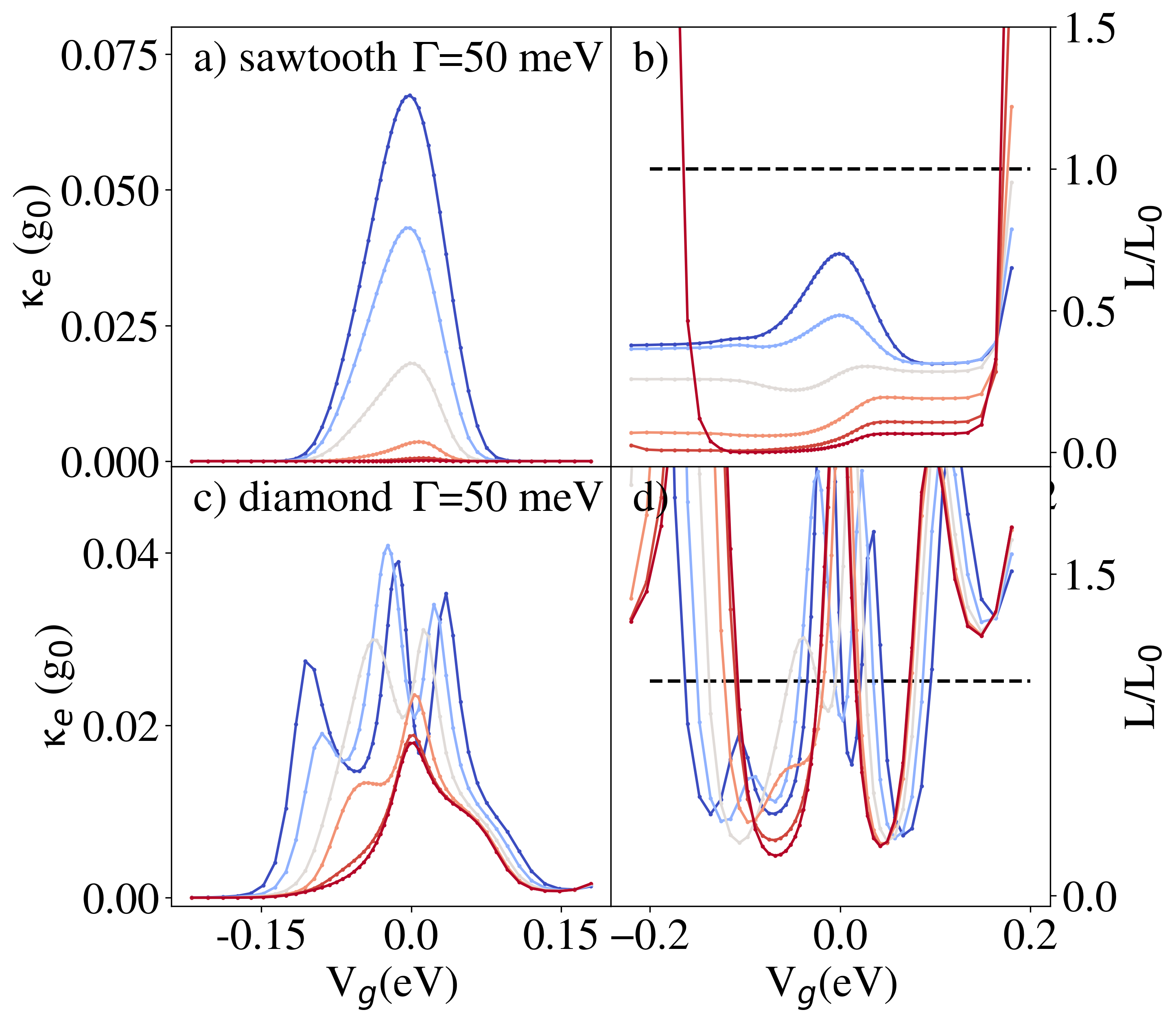}
\caption{(Color online). Electronic thermal conductivity \(\kappa_e\) in units of \(g_0\) (panels a), c)) and Lorenz ratio \(L/L_0\) normalized to the Sommerfeld value \(L_0\) (panels b), d)) for the sawtooth (top panels) and diamond (bottom panels) chains, computed at the Hartree-Fock level for \(\Gamma = 50\) meV, \(U = 0.15\) eV and \(T = 80\) K. Each curve corresponds to a different value of the bare non-interacting bandwidth \(W_n^0\), ranging from the wide-band (blue) to the narrow-band (dark red) limit, with the same values as in Fig.~\ref{fig:bandwidths}.}
\label{fig:eq_th_conductivity}
\end{figure}

Of particular interest is the violation of the Wiedemann-Franz (WF) law when the chemical potential falls within the narrow-band region. The WF law states that the ratio of the electronic thermal conductivity to the electrical conductivity is proportional to temperature, \(\kappa/\sigma = L_0 T\), through a proportionality constant that is largely universal across materials. This constant, the Lorenz number, takes the value \(L_0 = \pi^2/3\,(k_B/e)^2\), which in natural units (\(k_B = 1\), \(e = 1\)) gives \(L_0 \approx 3.29\). Physically, the WF law quantifies the heat carried by charge carriers, and deviations from it have been reported in a variety of systems~\cite{Kubala2008,Dutta2017,Majidi2022}. Although the microscopic origin of such violations is system-dependent, a breakdown of the WF law generically signals a decoupling between heat and charge currents.

A common extension of the WF law takes the form \(\kappa/\sigma = L\,T\), where \(L\) is the more general Lorenz number discussed in Sec.~\ref{sec:thermoelectric}. The ratio \(\eta = L/L_0\) then provides a quantitative measure of the WF law violation. Since the WF law is typically derived within the free-electron model for metals, values of \(\eta < 1\) (or \(\eta > 1\)) indicate that the conducting electrons carry less (or more) energy than their kinetic energy alone would predict.

In quantum systems where discrete energy levels can be individually resolved, such as quantum dots, violations of the WF law are ubiquitous~\cite{Kubala2008, Dutta2017, Karki2020, Majidi2022}. Consistently, WF law violations are found even in the non-interacting limit at low temperatures, an effect that weakens progressively as temperature rises. This can be understood intuitively: at very low temperatures, transport is dominated by electrons drawn from the leads at a well-defined energy, which disrupts the conventional relationship between electrical current and the energy transported by those electrons.

Panels b) and d) of Fig.~\ref{fig:eq_th_conductivity} display \(L/L_0\), as a function of the gate potential \(V_g\), with the dashed horizontal line marking \(L/L_0 = 1\). Here, the calculations are done at the Hartree-Fock level. In the sawtooth chain, panel b), \(L/L_0\) remains systematically below \(1\) across the entire gate voltage range, with the suppression becoming more pronounced as \(W_n^0\) is reduced. This plateau-like behavior over a finite gate voltage window is a direct consequence of the energy-filtering mechanism associated with the isolated flat-band, which suppresses energy fluctuations of the transmitted electrons. In the diamond chain, panel d), the behavior is richer and non-monotonic: \(L/L_0\) exceeds \(1\) in the vicinity of the flat-band, where the dispersive and flat-band transport channels contribute differently to heat and charge transport, before dropping sharply as the gate voltage crosses the flat-band. This resembles the filling dependence observed in a single-level quantum dot, with the minimum satisfying \(L < L_0\) near half-filling \cite{Dutta2017}, but the proximity of the dispersive band quickly restores transport and amplifies energy fluctuations, driving \(L/L_0\) back above unity.

However, neither of these signatures constitutes a physically meaningful departure from the Wiedemann-Franz law in the strict flatband limit. In the sawtooth chain, the suppression of \(L/L_0\) simply reflects the fact that both \(\kappa_e\) and \(\sigma\) vanish simultaneously as \(W_n^0 \to 0\), with \(\kappa_e\) vanishing faster, a ratio of two quantities both going to zero carries no useful thermodynamic information. In the diamond chain, the non-monotonic behavior of \(L/L_0\) is a consequence of the interplay between the two transport channels rather than a sign of genuine heat-charge decoupling. We have checked (not shown) that including many-body interactions at the GW level does not alter this picture qualitatively, though it does introduce quantitative corrections. Taken together, these observations suggest that while an isolated flatband appears to yield a more favorable Lorenz ratio, this advantage is fictitious: it does not translate into enhanced thermoelectric performance, as we now demonstrate through the analysis of the figure of merit \(zT\) in the following section.

\subsection{Seebeck coefficient}

\begin{figure}[t]
\includegraphics[width=\linewidth]{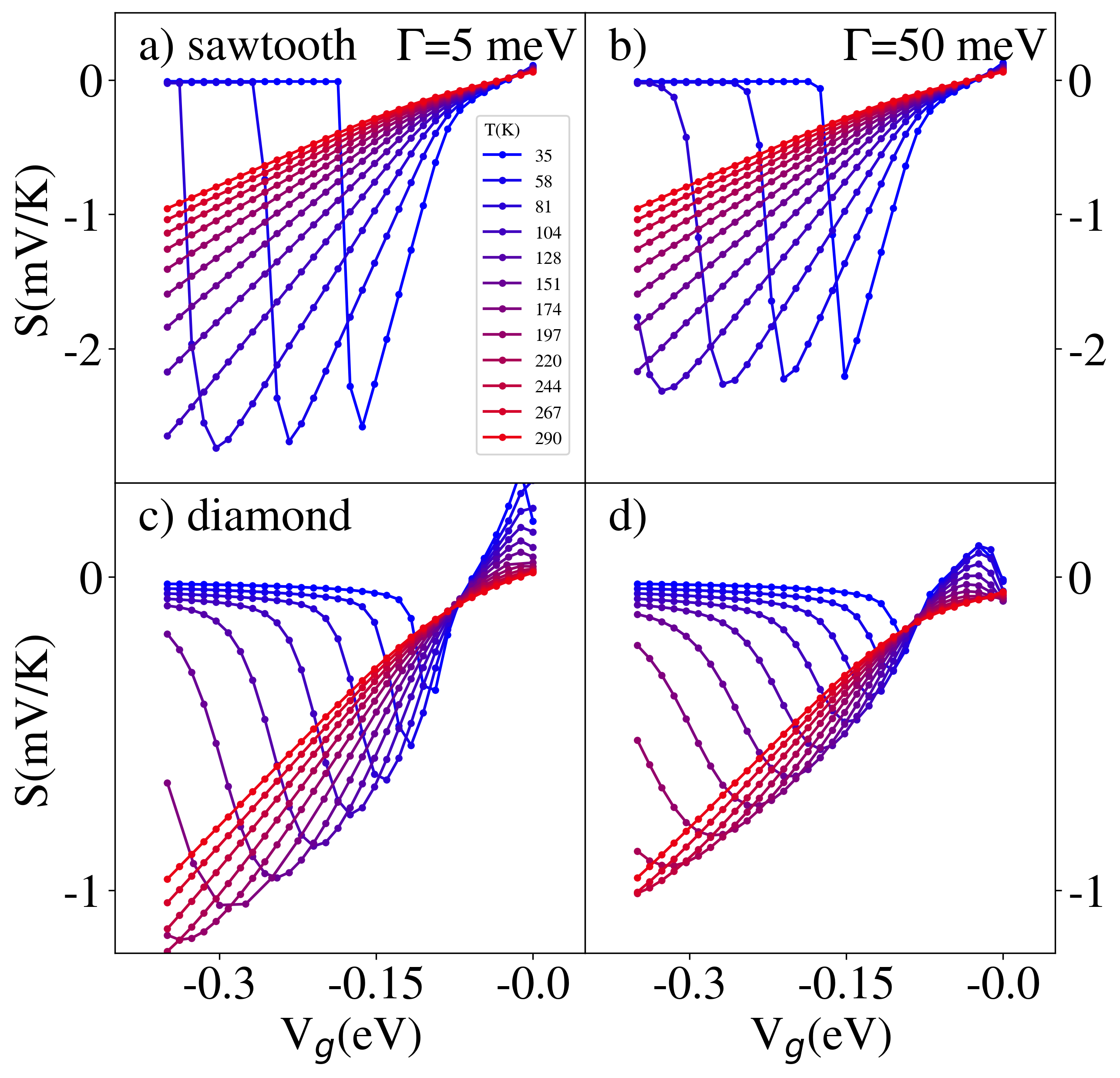}
\caption{(Color online). Seebeck coefficient \(S\) in units of mV/K as a function of the gate potential \(V_g\), computed at the Hartree-Fock level for \(U = 0.15\) eV in the region below half-filling of the narrowband, for the sawtooth chain (top panels, \(W_n^0 = 6.67\) meV) and the diamond chain (bottom panels, \(W_n^0 = 5.0\) meV), and for lead broadenings \(\Gamma = 5\) meV (left column) and \(\Gamma = 50\) meV (right column). Each curve corresponds to a different temperature ranging from \(T = 35\) K (dark purple) to \(T = 290\) K (dark red), as indicated in the legend.}
\label{fig:seebeck}
\end{figure}
The ability to convert temperature gradients into electric potential differences is quantified by the Seebeck coefficient. According to Eq.~\eqref{eq:seebeck}, a highly asymmetric transmission function around the chemical potential of the leads is sufficient to ensure a large Seebeck coefficient. This asymmetry typically arises from particle-hole asymmetry in the density of states, but can also be engineered through the coupling to the leads. In the case of flat-bands, the asymmetry originates from the sharpness of the density of states, which selects electrons from the leads at specific energies. A useful interpretation of Eq.~\eqref{eq:seebeck} is that the Seebeck coefficient measures the ratio between the average energy transported by electrons through the system \(\mathcal{I}_1\) and the thermal energy of the electrons in the leads \(T\sigma\). For a broad transmission function, as in metals, electrons transport on average exactly the thermal energy of the hottest lead, resulting in a small Seebeck coefficient. The sharpness of the flatband-induced transmission peak breaks this averaging and is the key mechanism behind the large values of \(S\) reported below.

Fig.~\ref{fig:seebeck} shows the Seebeck coefficient for the sawtooth chain, panels a) and b), and the diamond chain, panels c) and d), as a function of the gate potential \(V_g\), for lead-system couplings \(\Gamma = 5\) meV (left column) and \(\Gamma = 50\) meV (right column), and for temperatures ranging from \(T = 35\) K to \(T = 290\) K. In all panels \(S\) is negative throughout the explored gate range, consistent with electron-dominated transport below half-filling of the narrow-band. The overall magnitude of \(S\) reaches values of order \(1\)--\(2\) mV/K, which are exceptionally large compared to both conventional metallic systems and the best known thermoelectric materials \cite{Garmroudi2025a,Garmroudi2025b}, and directly reflect the sharpness of the narrow-band-induced peak in the transmission function.

In the weak-coupling regime (\(\Gamma = 5\) meV), panels a) and c), \(S\) develops a pronounced negative peak whose position shifts toward more negative gate voltages and whose magnitude grows with decreasing temperature, reflecting the increasingly sharp energy filtering provided by the flat-band edge at low \(T\). The sharp vertical drop visible in each curve marks the gate voltage at which the chemical potential enters the flat-band, where the electrical conductivity is suppressed and \(S\) diverges before becoming ill-defined. As temperature increases, the sharp feature is thermally smeared: the peak broadens and its magnitude is reduced, as the wider thermal window averages over a larger energy range of the transmission function, partially canceling the asymmetric contributions from either side of the narrowband peak. This temperature dependence is a direct manifestation of the competition between the energy selectivity of the narrowband and the thermal broadening of the Fermi distribution.

In the strong-coupling regime (\(\Gamma = 50\) meV), panels b) and d), the flatband peak in the transmission function is broadened by hybridization with the leads. As a consequence, the sharp dip in \(S\) is significantly reduced in depth and the overall gate-voltage dependence becomes smoother. Nevertheless, sizable absolute values of \(S\) persist even at large coupling, indicating that the narrowband still provides a substantial asymmetry in the transmission function even when hybridization is non-negligible. Comparing the two models, the sawtooth and diamond chains show qualitatively similar behavior, though the diamond chain displays a more symmetric gate-voltage profile around the narrowband energy, reflecting differences in the underlying band structure and the particle-hole symmetry properties of the two lattices.

We have demonstrated that flatbands are a powerful resource for thermoelectric energy conversion: the sharply peaked density of states associated with the flatband generates a strongly asymmetric transmission function, which is the key ingredient for a large Seebeck coefficient. The enhancement is most pronounced at low temperatures and weak coupling, where the energy selectivity of the flatband is least spoiled by thermal or hybridization broadening. Again, we have checked that treating many-body interactions at the GW level does not alter this picture qualitatively, but introduces moderate quantitative corrections to the values of \(S\). We present a comparative simulation of HF and GW in the case of the thermoelectric figure of merit analyzed next.

\subsection{Thermoelectric figure of merit $zT$}

\begin{figure}[t]
\includegraphics[width=\linewidth]{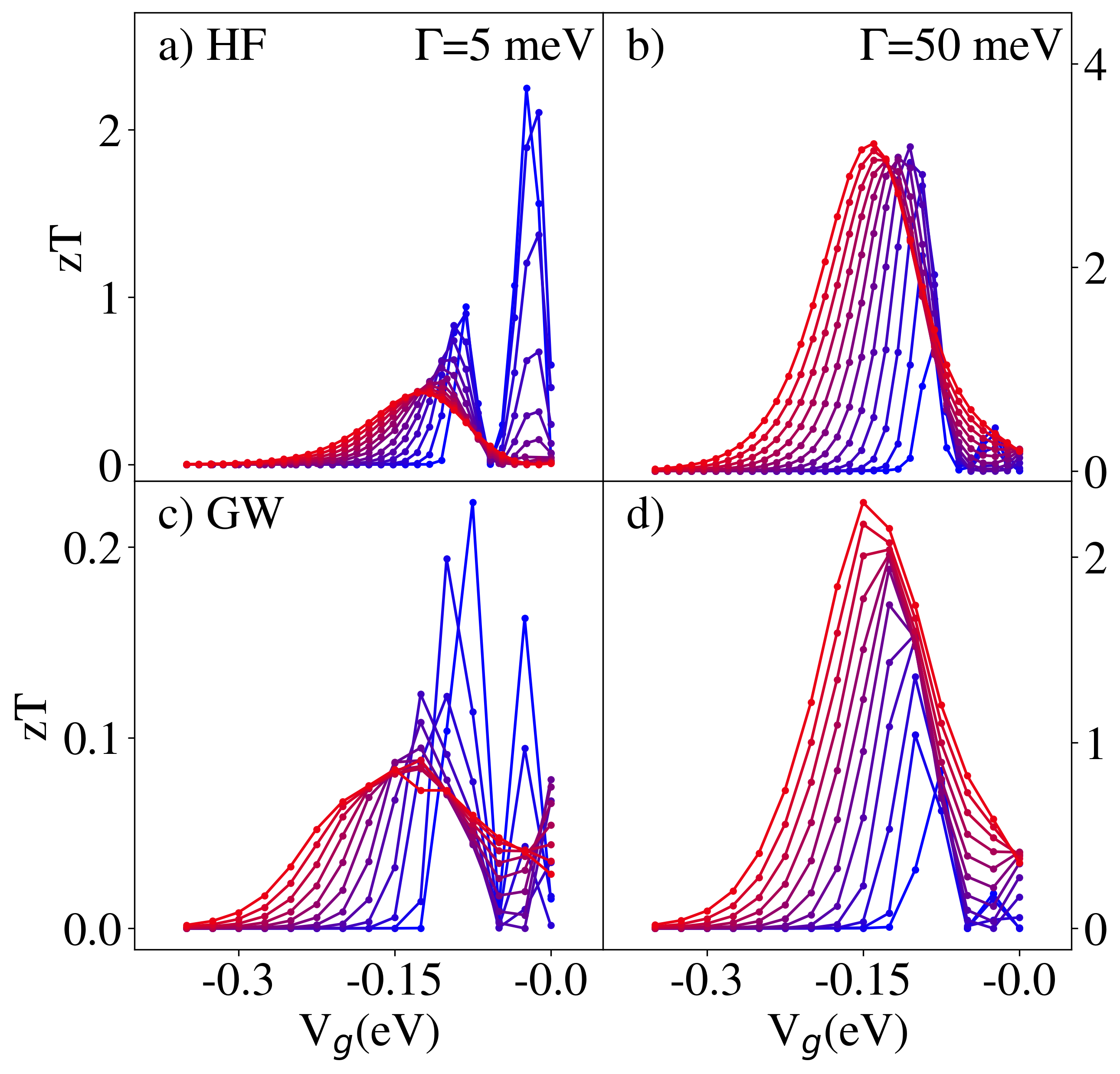}
\caption{(Color online). Thermoelectric figure of merit \(zT\) as a function of the gate potential \(V_g\) in the region below half-filling of the narrowband, computed for the diamond chain (\(W_n^0 = 5.0\) meV) for temperatures ranging from \(T = 35\) K (blue) to \(T = 290\) K (dark red), at the Hartree-Fock level (panels a) and b)) and at the GW level (panels c) and d)), for lead broadenings \(\Gamma = 5\) meV (left column) and \(\Gamma = 50\) meV (right column). The phonon thermal conductivity has been set to \(\kappa_{ph} = 10^{-3}g_0 T\) W/K, where \(g_0 T = \pi^2 k_B^2 T/3h\) is the quantum of thermal conductance. As discussed in the text, this value is negligible compared to the electronic thermal conductivity in the wide-bandwidth regime, but becomes comparable to \(\kappa_e\) as the flatband limit is approached, so that the values of \(zT\) shown here represent an upper bound only for the largest bandwidths.
}
\label{fig:zt}
\end{figure} 

Both the Lorenz ratio and the Seebeck coefficient suggest that an isolated flatband outperforms a flatband embedded in a dispersive background. This stands in apparent tension with the picture put forward in \cite{Garmroudi2025a,Garmroudi2025b}, where thermoelectric enhancement is attributed to the restoration of flatband transport via scattering into a dispersive band. However, a full restoration of transport would necessarily suppress the Seebeck coefficient, driving the material toward conventional metallic behavior — which is precisely the opposite of what is desired.

A crucial caveat must be kept in mind here. Although the sawtooth model yields a large thermovoltage, its electrical conductivity vanishes, as shown in Fig.~\ref{fig:max_eq_el_conductivity}. A large Seebeck coefficient under these conditions does not translate into useful thermoelectric performance. The same reasoning applies to the Lorenz ratio: its anomalously low value does not reflect a genuinely favorable departure from the Wiedemann-Franz law, but rather the fact that the thermal conductivity vanishes even faster than the electrical conductivity. The apparent violation of the Wiedemann-Franz law is therefore physically meaningless in this limit. To make this concrete, we now turn to the thermoelectric figure of merit, \(zT\), as defined in Eq.~\eqref{eq:zt}, which simultaneously accounts for all three transport coefficients and provides an unambiguous measure of thermoelectric performance. 

Fig.~\ref{fig:zt} shows \(zT\) as a function of the gate potential \(V_g\) in the region below half-filling of the narrowband, for temperatures ranging from \(35\) K (blue) to \(290\) K (dark red), computed with \(\kappa_{ph} = 10^{-3} g_0\) W/K, where \(g_0 = \pi^2 k_B^2 T / 3h\) is the quantum of thermal conductance. As can be seen from Fig.~\ref{fig:eq_th_conductivity}, this is not universally negligible: while \(\kappa_{ph}\) is small compared to the electronic thermal conductivity at large \(W_n^0\), it becomes comparable to or larger than \(\kappa_e\) as the flatband limit is approached and the electronic conductivity is suppressed. The values of \(zT\) shown in Fig.~\ref{fig:zt} therefore represent a true upper bound only in the wide-bandwidth regime, while in the narrow-bandwidth limit the phonon contribution already plays a non-trivial role in limiting thermoelectric performance. This is the case for the chosen \(W_n^0\) for both models. A more systematic treatment of the phononic thermal conductivity goes beyond the scope of our work and requires a proper definition of the physical system; the values we consider here illustrate the role of \(\kappa_{ph}\) in determining the upper bound for \(zT\).
All four panels display a pronounced peak near \(V_g \approx -0.15\) eV, which corresponds to the chemical potential being tuned to the vicinity of the flatband edge where the Seebeck coefficient is largest. It is important to note that the narrowband itself is located in the energy window \(-0.08\) eV \(\lesssim V_g \lesssim -0.02\) eV within the gate range explored here. The peak in \(zT\) therefore does not occur when the chemical potential sits inside the narrowband, but rather just below it, in the energy region where the density of states varies most rapidly with energy. This is fully consistent with the Mahan-Sofo picture~\cite{Mahan1996}: the optimal thermoelectric response is achieved not at the center of a sharp spectral feature, but at its lower edge, where the energy derivative of the transmission function is maximized and the Seebeck coefficient consequently peaks. As the gate potential is swept through the narrowband region, \(zT\) drops sharply: once the chemical potential enters the flatband, conductivity is suppressed and the system enters the paradoxical regime discussed above, where a large thermovoltage coexists with vanishing carrier mobility.

At the Hartree-Fock level, panels a) and b), the peak values of \(zT\) are remarkably large for both values of the lead broadening, with the larger \(\Gamma = 50\) meV yielding superior performance consistent with our earlier conclusion that a finite hybridization with the dispersive background is necessary to restore carrier mobility without fully suppressing the Seebeck coefficient. The \(zT\) peak grows and sharpens with increasing temperature, reflecting the growing importance of thermally activated transport across the flatband edge. At the GW level, panels c) and d), correlation effects beyond mean field substantially reduce the peak values of \(zT\), indicating a significant renormalization of the thermoelectric response through a redistribution of spectral weight and an enhancement of effective scattering rates. Notably, the secondary sharp feature visible near \(V_g \approx -0.05\) eV in panels a) and c) at low temperatures and small broadening can be directly associated with transport through the narrowband itself: its suppression at larger \(\Gamma\) and higher \(T\) confirms that it originates from fine structure in the transmission function tied to the flatband density of states, which is progressively smeared out by interaction-induced, hybridization and thermal broadening.

\section{Conclusions}

In this work we have investigated the thermoelectric properties of interacting electrons in flatband systems, using two paradigmatic one-dimensional models, the sawtooth chain and the diamond chain, as representative cases of an isolated flatband and a gapless flatband touching a dispersive band, respectively. The two models were treated within a non-equilibrium Green function framework in which electron-electron interactions are included diagrammatically at the Hartree-Fock and GW levels of approximation, allowing us to access interaction-driven renormalizations of both the spectral function and the transport coefficients self-consistently and beyond the mean-field description.

A central result of this work is that the naive expectation, that a perfectly flat band should be an optimal thermoelectric, is physically flawed. Although the Seebeck coefficient is large and the Lorenz ratio anomalously suppressed in the strict flat-band limit, these signatures alone are meaningless: the electrical conductivity vanishes as the chemical potential enters the flat-band, so that a large thermovoltage is generated but no current can flow, and the apparent violation of the Wiedemann-Franz law simply reflects the fact that thermal conductivity vanishes faster than electrical conductivity. This connects directly to the result of Mahan and Sofo~\cite{Mahan1996}: a delta-function transmission is mathematically optimal but physically unreachable, since a perfectly flat-band implies fully quenched kinetic energy and a vanishing transmission function. Some finite broadening, whether from hybridization with dispersive bands introduced by the leads or from scattering into nearby dispersive states, is not a perturbation to be minimized but a necessary condition for useful thermoelectric performance. This clarifies the apparent disagreement with the scenario proposed in \cite{Garmroudi2025a,Garmroudi2025b}: a partial restoration of flat-band transport is beneficial, but a full restoration suppresses the Seebeck coefficient and drives the system toward conventional metallic behavior.

The thermoelectric figure of merit, $zT$, confirms this picture: its peak occurs just below the flat-band edge, where the transmission function varies most rapidly with energy, rather than inside the flat-band itself. The systematic comparison between HF and GW results reveals that mean-field treatments generally overestimate thermoelectric performance, and that a proper account of electronic correlations is necessary for reliable predictions.
These findings provide a coherent and physically transparent set of design principles for flatband-based thermoelectric devices, and a powerful framework for their theoretical description using the NEGFs approach. Natural extensions include two-dimensional flat-band materials such as twisted moiré systems, kagome metals, and other strongly correlated platforms where both interaction effects and thermoelectric functionality are of active experimental interest.

\begin{acknowledgments}
Numerical simulations were performed exploiting the Finnish CSC facilities under the Project no. 2009128 (``Transport in flatband materials'').
R.T. acknowledges the financial support of the Jane and Aatos Erkko Foundation (Project EffQSim) and the Research Council of Finland through the Finnish Quantum Flagship (Project No. 359240).
\end{acknowledgments}
\appendix
\section{Physical quantities within the NEGFs approach}
\label{app:negfs}

To compute the physical quantities of interest we use the non-equilibrium Green functions (NEGFs)~\cite{Ridley2022} approach, which allows us to treat the leads non-perturbatively and the many-body interaction through perturbation theory.
For this reason, NEGFs are particularly suitable for studying transport~\cite{Talarico2019,Talarico2020,Ridley2022} in correlated systems. 
Moreover in the case of non-interacting systems the quantities of interests such as currents and conductivities are exact unlike in other approaches which use linear response which is basically a perturbation approach in the coupling strengths to the leads.

We study the non-equilibrium steady state and specifically the electric current and the conductivity. We solve the Dyson equation in the frequency domain which reads:
\begin{align}
\label{eq:ret_green}
G^R(\omega)&=g^R(\omega)+g^R(\omega)\Sigma^R(\omega)G^R(\omega)\\
\label{eq:lesser_green}
G^<(\omega)&=G^R(\omega)\Sigma^<(\omega)G^A(\omega)\\
\Sigma^{\text{X}}(\omega)&=\Sigma^{\text{X}}_{\text{HY}}(\omega)+\Sigma^{\text{X}}_{\text{MB}}(\omega)
\end{align}
where $g$ is the reference Green function, which we take as the Hartree-Fock one.
Above, \(\text{X}=<,R,A\) and the \(G^{\text{X}}(\omega)\) are the Fourier transforms of the corresponding single particle Green functions:
\begin{align}
\mathcal{G}^R_{\mathbf{x}\mathbf{x}'}(t-t')&=-\text{i}\;\theta(t-t')\left\langle{\left[\hat{c}_{\mathbf{x}}(t),\hat{c}_{\mathbf{x}'}(t')^\dagger\right]_{+}}\right\rangle_0\\
\mathcal{G}_{\mathbf{x}\mathbf{x}'}^<(t-t')&=\text{i}\left\langle{\hat{c}_{\mathbf{x}'}^\dagger(t')\hat{c}_{\mathbf{x}}(t)}\right\rangle_0,
\end{align}
where we introduced a multi-index \(\mathbf{x}=(i,\alpha,\sigma)\).
Here the average is taken over the initial equilibrium states of the uncoupled non-interacting systems.
An important assumption is that stationarity is reached and therefore the single particle Green function depend only on the relative time \(\tau=t-t'\), e.g., $G^<(\omega)=\int d\tau e^{i\omega\tau} \mathcal{G}^<(\tau)$.
The total self-energy is the sum of the lead induced hybridization self-energies \(\Sigma^{\text{X}}_{\text{HY}}(\omega)=\sum_\alpha\Sigma^{\text{X}}_{\text{HY},\alpha}(\omega)\) accounting for the coupling with the leads \(\alpha=L,R\) and the many-body self-energy \(\Sigma^{\text{X}}_{\text{MB}}(\omega)\) accounting for the electron-electron scattering.
The components of the hybridization self-energies are given by:
\begin{align}
\Sigma^{R}_{\text{HY},\alpha}(\omega)&=-\text{i}\frac{{\bf\Gamma}_\alpha}{2}\\
\Sigma^{<}_{\text{HY},\alpha}(\omega)&=\text{i}{\bf\Gamma}_\alpha\;f_\alpha(\beta_\alpha(\omega-\mu_\alpha))
\end{align}
with \(f(x)=\left(e^x+1\right)^{-1}\) the Fermi-Dirac distribution.
These self-energies are obtained in the wide-band limit, namely the density of states of the leads is constant over the range of the density of states of the system we are interested in.
The matrices \({\bf \Gamma}_\alpha\) carry the information about the
coupling strength between the lead \(\alpha\) and the system as well as 
information on the geometry of the coupling.
We take the following form for the coupling matrices
\begin{align}
\left({\bf \Gamma}_{G}\right)_{\bf{x}\bf{x}'}&=\Gamma_G\delta_{\sigma\sigma'}\delta_{i,j}\delta_{\alpha,\alpha'}\\
\left({\bf \Gamma}_{L}\right)_{\bf{x}\bf{x}'}&=\Gamma_L\delta_{\sigma\sigma'}\delta_{i,1}\delta_{i',1}\gamma_{\alpha,\alpha'}
\nonumber\\
\left({\bf \Gamma}_{R}\right)_{\bf{x}\bf{x}'}&=\Gamma_R\delta_{\sigma\sigma'}\delta_{i,N}\delta_{i',N}\gamma_{\alpha,\alpha'},
\nonumber
\end{align}
where the matrices \(\gamma_{\alpha,\alpha'}\) are given by:
\begin{equation}
\gamma_{\alpha,\alpha'}=\sum\limits_{s,s'}\delta_{\alpha,s}\delta_{\alpha',s'},
\end{equation}
with the sums running over \(a,b\) for the sawtooth and \(a,b,c\) for the diamond.
With this geometry the driving leads \(L,R\) are coupled to the first and last cell of the system respectively.

We refer the reader to the Refs~\cite{LoGullo2016,Talarico2019,Talarico2020,Ridley2022} for the details of the derivation of the equations above. For the many-body self-energies we consider the GW approximations as discussed in the main text.
Through the single-particle Green functions in Eqs.~\eqref{eq:ret_green}
and \eqref{eq:lesser_green} it is possible to extract information about physical quantities of interests.
In particular, the total current through the system is defined as \(I=I_{L}-I_{R}\) with \(I_{\alpha}\equiv -e\;d\langle\hat{N}_{\alpha}\rangle/dt\), where \(\hat{N}_{\alpha}\) is the total particle number operator of the lead \(\alpha\). The current in the left lead is given by: 
\begin{equation}
\label{appeq:el_curr}
I_{L}=\frac{e}{\hbar}\int\limits_{-\infty}^{\infty}\frac{d\omega}{2\pi}\;\mathcal{T}(\omega) (f(\beta_L(\omega-\mu_L))-f(\beta_R(\omega-\mu_R))),
\end{equation}
where the transmission function is \(\mathcal{T}(\omega)=\text{Tr}(\Gamma_{L}G^R(\omega)\Gamma_{R}G^A(\omega))\). A similar equation holds for the current flowing to the right lead.

Analogously, the energy current is defined as the variation of energy of the lead \(J_{\alpha}\equiv\;d\langle\hat{H}_{\alpha}\rangle/dt\) and which can be easily shown to be:
\begin{equation}
\label{appeq:he_curr}
J_{L}=\frac{1}{\hbar}\int\limits_{-\infty}^{\infty}\frac{d\omega}{2\pi}\;\omega\mathcal{T}(\omega) (f(\beta_L(\omega-\mu_L))-f_R(\beta_R(\omega-\mu_R))).
\end{equation}
It is worth mentioning that at stationarity we have, by definition, that the number of particles in the system does not change and therefore \(I_L=-I_R\) which follows from the conservation of the total number of particles. Therefore we have \(I=2I_L\), and for this reason we consider only \(I_L\). The full expression \(I=I_{L}-I_{R}\) is useful in the transient when the two currents need not to be equal in value and opposite in sign. The same hold for the energy currents, and therefore for the heat currents defined as \(\dot Q_\alpha=J_\alpha-\mu_\alpha I_\alpha/e\).

From the current we compute the electrical and thermal conductivities of the system by expanding to first order in both \(\Delta V\) and \(\Delta T\):
\begin{eqnarray}
\label{appeq:linear}
(f(\beta_L(\omega-\mu_L))-f_R(\beta_R(\omega-\mu_R)))\\\approx \frac{\partial f}{\partial \omega}\left(-e\Delta V- \frac{(\omega-\mu)}{T}\Delta T\right),\nonumber
\end{eqnarray}
where \(T\) is the temperature. This gives
\begin{align}
    \sigma&=\frac{e^2}{\hbar}\int\limits_{-\infty}^{\infty}\frac{d\omega}{2\pi}\;\mathcal{T}(\omega) \left(-\frac{\partial f}{\partial \omega}\right), \label{appeq:el_cond}\\
    \kappa_0&=\frac{1}{T \;\hbar}\int\limits_{-\infty}^{\infty}\frac{d\omega}{2\pi}\;\mathcal{T}(\omega) \left(-\frac{\partial f}{\partial \omega}\right)(\omega-\mu)^2.\label{appeq:th_cond}
\end{align}
Inserting Eq.~\eqref{appeq:linear} in Eq.~\eqref{appeq:el_curr} it is possible to derive an expression for the thermovoltage \(\Delta V_{th}\) such that \(I_L(\Delta V_{th},\Delta T)=0\) and for the Seebeck coefficient:
\begin{equation}
    \Delta V_{th}=-\frac{e}{\sigma T}\Delta T
    S=-\frac{\Delta V_{th}}{\Delta T}=\frac{e}{\sigma T}.
\end{equation}

\bibliographystyle{plain}
\bibliography{thermoelectric}

\end{document}